\documentclass[%
 reprint,
superscriptaddress,
 nofootinbib,
 amsmath,amssymb,
 aps,
 prd,
]{revtex4-2}

\usepackage{xcolor}
\usepackage{multirow}
\usepackage{graphicx}
\usepackage{dcolumn}
\usepackage{bm}
\usepackage{acronym}
\usepackage{hyperref}
\usepackage{orcidlink}
\usepackage{array}
\usepackage{subcaption}
\usepackage[table]{xcolor}
\usepackage{ulem}

\definecolor{garibaldired}{HTML}{DD0000}

\newcommand{\updated}[1]{#1}
\newcommand{\deleted}[1]{}

\newcommand{\strutrow}{\rule{0pt}{2.5ex}} 
\newcommand{\portsmouth}{Institute of Cosmology \& Gravitation, University of Portsmouth, Portsmouth, United Kingdom}

\begin{document}

\preprint{APS/123-QED}

\title{Inpainting to search for premerger massive black hole binaries in realistic LISA datasets.}

\author{Gareth {Cabourn Davies}\orcidlink{0000-0002-4289-3439}}
\email{gareth.cabourn-davies@port.ac.uk}
\affiliation{\portsmouth}
\author{Ian Harry\orcidlink{0000-0002-5304-9372}}
\email{ian.harry@port.ac.uk}
\affiliation{\portsmouth}

\begin{abstract}
    A key science target of the Laser Interferometer Space Antenna (LISA) is to carry out
    multi-messenger observations of massive black hole binaries, observing the merger
    simultaneously in gravitational waves and with electromagnetic observatories.
    Identifying that a merger is happening and providing an updating estimate of the sky location in the hours, days and weeks before the merger is critical to enable electromagnetic observations of the merger event. 
    In this work we demonstrate and compare two methods for premerger identification of massive
    black hole binaries; a zero-latency filter approach and, for the first time, an approach using an ``inpainting'' technique.
    We apply these methods to the LISA Data Challenge dataset 2a--Sangria-HM--and demonstrate the
    successful recovery of the 14 signals in the dataset that we expected to be identifiable at
    least half a day before merger.
    We demonstrate that the inpainting method can identify premerger signals even when gaps are present in the data,
    demonstrating the recovery of a signal even when three day-long data gaps are added to the 14 days preceding merger.
    Finally, we explore the challenge of overlapping signals, using a region of overlapping signals in the Sangria-HM dataset, all of which merge within a 10-day window, and show how removing signals that have been confidently identified from the data allows us to identify quieter signals in the same period.
\end{abstract}

\maketitle

\begingroup
    \renewcommand\thefootnote{}
    \footnotetext{Both authors contributed equally to this work.}
    \addtocounter{footnote}{-1}
\endgroup

\acrodef{LISA}[LISA]{Laser Interferometer Space Antenna}
\acrodef{MBHB}[MBHB]{Massive Black Hole Binary}
\acrodefplural{MBHB}[MBHBs]{Massive Black Hole Binaries}

\section{Introduction}
\label{sec:intro}

The \ac{LISA} gravitational-wave observatory, planned for launch in the mid-to-late
2030s~\cite{LISA:2017pwj, Colpi:2024xhw}, will unlock our ability to observe the universe
in the low-frequency gravitational-wave band~\cite{LISA:2022yao}. A key science target
for \ac{LISA} is the observation of \acp{MBHB}. These signals are expected to have a very large
signal-to-noise ratio, making the problem of identifying them trivial~\cite{LISA:2022yao}.
A significant challenge lies in reliably inferring the parameters of these sources, especially
given a large population of other sources present in the data, and this problem has been
extensively explored over the last two decades, primarily through the LISA Data
Challenges~\cite{Arnaud:2006gm, MockLISADataChallengeTaskForce:2006sgi, Arnaud:2007vr,
MockLISADataChallengeTaskForce:2007iof, Babak:2008aa, Arnaud:2007jy,
MockLISADataChallengeTaskForce:2009wir, Baghi:2022ucj, LDC_WEBSITE}.

The large signal-to-noise ratio that \acp{MBHB} are expected to have unlocks a particularly
interesting possibility; that of observing these systems
in the weeks to months before they merge~\cite{CabournDavies:2024hea, Houba:2024mqj,
Ruan:2024qch}. Such \textit{pre-merger} observations provide an early warning that a loud merger
event is about to happen and can provide sky-localization
estimates~\cite{CabournDavies:2024hea, Kocsis:2007yu, McWilliams:2011zs, Saini:2022hrs}
for these incredibly science-rich events~\cite{DalCanton:2019wsr}.
This advanced notice is critical to enable
electromagnetic facilities to be directed toward the source region rapidly to maximize
the possibility of multi-messenger observation and to potentially reschedule any planned repositioning operations of \ac{LISA} itself 
to reduce the chance that it is not taking data when the merger arrives~\cite{Colpi:2024xhw,Castelli:2024sdb}.

The technique that we introduced in~\cite{CabournDavies:2024hea} offers a method for immediate analysis of LISA data.
It allows for the accurate computation of a matched filter using an acausal, zero-latency whitening filter that alleviates the need to wait (up to a day) for additional data to arrive before being able to filter.
If one has to wait an additional day, the observation window may have passed, or \ac{LISA} itself may not be taking data while repositioning~\cite{Colpi:2024xhw,Castelli:2024sdb}.

\begin{figure*}
    \includegraphics[width=\textwidth]{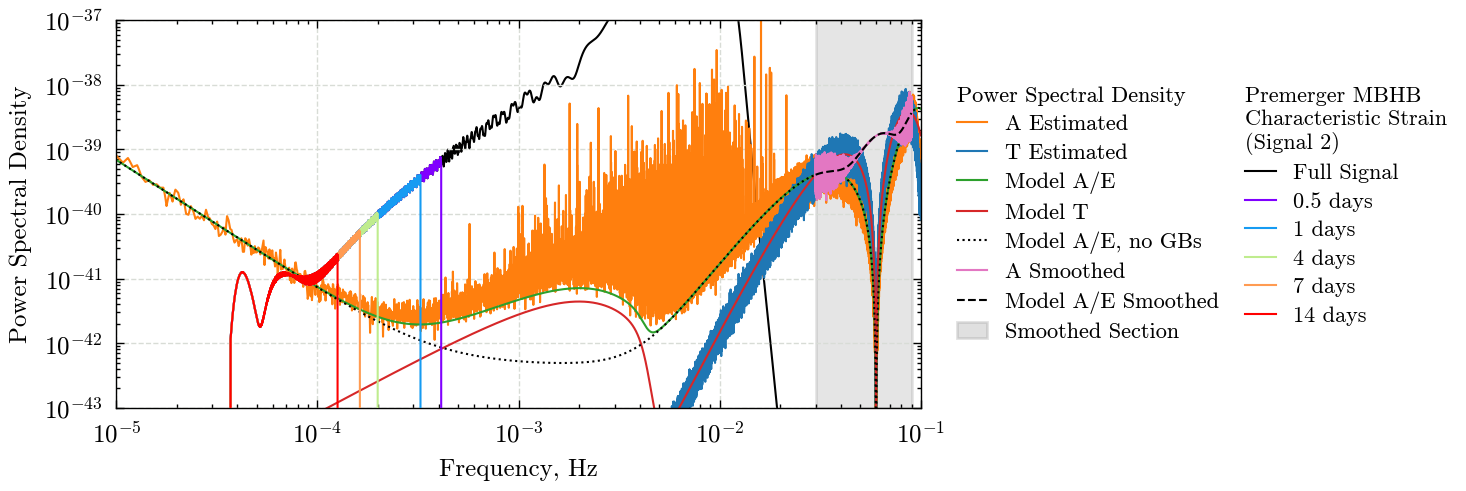}
    \caption{
        Comparison of power spectral densities used in this work.
        The power spectral density is estimated from the Sangria-HM dataset using the Welch method \updated{as described in~\cite{Allen:2005fk}}.
        \updated{We use the full Sangria dataset, median averaging across segments, applying Hann windowing, and using a segment duration of 18.25 hours and a stride of 9.125 days}.
        The model shown is the noise model used to produce the Sangria-HM dataset, and is shown with and without unresolvable Galactic binary sources.
        We show the power spectral density model and estimate smoothed using the method described in Section~\ref{subsec:smoothing}.
        We also show the characteristic strain for a representative \ac{MBHB} signal from Sangria-HM (Signal 2), which shows that at 14, 7 and 4 days-before-merger, the \ac{MBHB} does not overlap
        the Galactic binary confusion noise in frequency, but when the signal reaches around one day before merger (blue line),
        the Galactic binary confusion noise starts to overlap with the signal.
    }
    \label{fig:psds}
\end{figure*}

However, our method is not robust to gaps in the LISA data and only searches for signals at
discrete values of the gap to merger. We therefore introduce a second possibility for premerger
observation, applying the inpainting technique of~\cite{Zackay:2019kkv} as an alternative to the
zero-latency whitening filter.

In this paper, we present the application of both the zero-latency and inpainting techniques to the
``Sangria-HM dataset'' produced as part of the LISA Data Challenges.
We demonstrate that, where there are no gaps in the data, both methods
compare very well, and identify 14 of the 15 \ac{MBHB} signals in the dataset at least 0.5 days before merger--the remaining signal being too quiet for detection 0.5 days before merger.
We demonstrate how the inpainting technique works well when gaps are added to the dataset, allowing premerger identification even in an example where the data contains three separate day-long gaps.

The paper is organized as follows: In section~\ref{sec:dataset} we describe the Sangria-HM dataset and evaluate the potential for premerger detection of the 15 \ac{MBHB} signals added to it.

In section~\ref{sec:zerolatency_sangria} we present the results of our zero-latency search on this dataset.
In section~\ref{sec:inpainting} we describe how we adapt the inpainting technique of~\cite{Zackay:2019kkv} to the problem of premerger observation of \acp{MBHB} before presenting results of that search in section~\ref{sec:inpainting_sangria}.
In section~\ref{sec:focused} we discuss the analysis of Signals 2 - 5, whose merger times are all contained within a single ten-day window, causing some overlap in their identification.
Finally we conclude in section~\ref{sec:conclusion}.

All of our results and figures are fully reproducible, with the code
used to make them publicly available. Please see the data release at
\url{https://icg-gravwaves.github.io/lisa_premerger_sangria/}
for more detail.

\section{The Sangria-HM Dataset}
\label{sec:dataset}

The LISA Data Challenges were designed to simulate comprehensively modeled data as an introduction
to LISA data analysis~\cite{Baghi:2022ucj, LDC_WEBSITE}.
Here we use the dataset known as Challenge 2a, also known as
\textit{Sangria-HM}~\cite{LDC_WEBSITE_SANGRIA_HM}.
Sangria-HM  consists of both a ``training'' and ``blind'' dataset, we analyze here the training dataset to more easily compare the values of the signals to our results.
The Sangria-HM training dataset is a year long and contains Gaussian instrumental noise, a foreground of
Galactic binaries and \ac{MBHB} systems, both with parameters derived from astrophysical models.
The \acp{MBHB} are simulated using the \texttt{PhenomHM} waveform model~\cite{Garcia-Quiros:2020qpx}.

In our previous work~\cite{CabournDavies:2024hea}, we demonstrated the zero-latency technique
on a dataset that did not include a foreground of Galactic binaries.
This foreground, made up of millions of white dwarf binaries in our galaxy which are nearly monochromatic in frequency, occupy the same frequency range as mergers of \acp{MBHB}.
Many of the Galactic binary systems are not expected to be resolvable, and so must be considered an extra source of noise for \ac{MBHB} searches.

The contribution of this `noise' will vary throughout the year, as LISA's most sensitive direction is oriented toward or away from the galactic center~\cite{Colpi:2024xhw,Seto:2004ji}.
As the loudest Galactic binary sources are resolved and can be removed from the data, sensitivity to \ac{MBHB} signals will naturally increase~\cite{Crowder:2006eu,Strub:2024kbe}.

The Galactic binary confusion noise overlaps with \ac{MBHB} mergers in frequency space.
But we emphasize that our search is not \textit{for} the mergers,
rather the inspiral stage of the signal, many days before merger, and so we consider whether the frequency overlaps with the
particular times-before-merger for our signals.

Figure~\ref{fig:psds} shows the power-spectral density directly estimated from the data, the modeled power-spectral density with and without Galactic binary foreground noise, and an example
characteristic strain curve for Signal 2 in the Sangria-HM dataset (See Section~\ref{subsec:expected_snr} for a description).

From Figure~\ref{fig:psds}, we see that at 14, 7 and 4 days-before-merger, the pre-merger characteristic strain does not overlap
in frequency with the unresolvable Galactic binary confusion noise.
At 1 and 0.5 days before merger, the Galactic binary confusion noise begins to overlap the \ac{MBHB}.
However, the ``day 0'' sensitivity curve, where no resolvable Galactic binaries have been removed, qualitatively agrees with the modeled sensitivity curve where resolvable binaries are removed.
This is important because it indicates that \emph{premerger detection of \ac{MBHB} \deleted{is not a ``Global Fit'' problem}\updated{can be run immediately at the start of the mission}}.
The frequency content of premerger \acp{MBHB} are broadly separated from any other resolvable source, and therefore one would not need to simultaneously perform inference of premerger \acp{MBHB} and any other source because premerger inference could run perfectly well on data where no sources have been removed from it.
It also indicates that the techniques to observe \ac{MBHB} sources could largely run separately from Global
Fit analyses, where all sources are being simultaneously fit\footnote{As we will show in Section~\ref{subsec:peak_removal}, there is sufficient correlation between premerger \acp{MBHB} and \ac{MBHB} mergers that one would want to remove observed \acp{MBHB} from the data as they are observed to aid detection of potentially overlapping premerger \acp{MBHB}.}.
\updated{This point is explored further in Appendix~\ref{sec:gb_noise_snr_loss}.}

\begin{figure}
    \centering
    \includegraphics[width=\columnwidth]{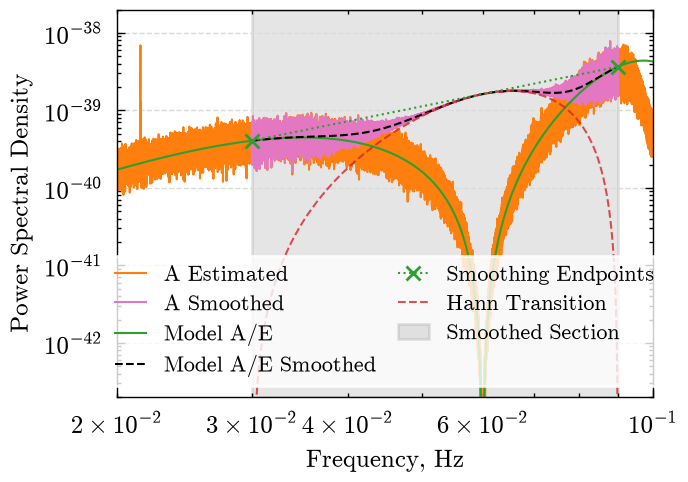}
    \caption{
        The method used to smooth over the dip in the power-spectral density at 60 mHz.
        We draw a straight line between the two peaks either side of this point, and smoothly
        transition between the estimated power-spectral density and the straight line using a Hann window.
        This ensures that there are no immediate jumps in the power-spectral density to cause spectral leakage.
        This is applied to both the estimated and modeled power-spectral density used in later analysis.
    }
    \label{fig:smoothing}
\end{figure}

\subsection{MBHB signals in the Sangria-HM data set}
\label{subsec:expected_snr}
For the example \ac{MBHB} signal shown in Figure~\ref{fig:psds}, we use the \textit{characteristic
strain}~\cite{Moore:2014lga} for Signal 2.
Characteristic strain has a desirable property for visualization; the integrated area between the characteristic strain and the
power-spectral density in the frequency domain corresponds to the expected signal-to-noise ratio for the signal.

\updated{Signal-to-noise ratios calculated from integrated areas below the} characteristic strains at the times-before-merger that we consider are shown in
Table~\ref{tab:characteristic_strain_snr}.
We compute this for each signal in the Sangria-HM dataset, with cutoffs set according to the
time--frequency evolution expected for an inspiralling \ac{MBHB} signal.
We see that Signal 1 is unlikely to be seen even 0.5 days pre-merger, and that only the loudest signals -
0, 3, 4, 10, 11 and possibly 12, 13 - would be seen 4 days or more before merger.

\begin{table}
    \begin{tabular}{|cc|c|ccccc|}
    \hline
    \multicolumn{2}{|c|}{Signal}& \multicolumn{6}{c|}{Expected SNR} \\
    \hline
    Number & Time  & Full Signal & 0.5 & 1 & 4 & 7 & 14 \\
    \hline
    0 & 55.56 & 2699.97 & 70.99 & 52.74 & 24.33 & 16.43 & 9.51 \\
    1 & 101.23 & 105.73 & 3.99 & 2.49 & 0.79 & 0.47 & 0.28 \\
    2 & 129.26 & 400.33 & 10.76 & 7.51 & 2.91 & 1.84 & 1.07 \\
    3 & 130.31 & 1992.40 & 71.92 & 54.76 & 27.52 & 19.88 & 13.22 \\
    4 & 133.41 & 3475.54 & 115.99 & 88.26 & 44.02 & 31.65 & 20.46 \\
    5 & 138.55 & 427.00 & 16.40 & 10.79 & 3.85 & 2.37 & 1.22 \\
    6 & 157.61 & 673.55 & 24.99 & 16.19 & 5.48 & 3.25 & 1.55 \\
    7 & 191.34 & 579.20 & 15.72 & 10.70 & 4.02 & 2.56 & 1.49 \\
    8 & 199.60 & 365.55 & 16.21 & 10.68 & 3.77 & 2.28 & 1.11 \\
    9 & 215.34 & 392.92 & 18.69 & 12.41 & 4.75 & 3.12 & 1.82 \\
    10 & 236.41 & 2189.05 & 45.09 & 31.90 & 12.71 & 8.13 & 4.37 \\
    11 & 257.27 & 1200.61 & 45.32 & 33.31 & 13.99 & 8.92 & 4.71 \\
    12 & 271.29 & 359.01 & 21.27 & 14.71 & 6.10 & 4.17 & 2.57 \\
    13 & 282.51 & 822.50 & 24.75 & 18.00 & 7.81 & 5.23 & 3.00 \\
    14 & 341.62 & 230.69 & 9.88 & 6.43 & 2.22 & 1.34 & 0.67 \\
    \hline
    \end{tabular}
    \caption{
        Optimal signal-to-noise ratios (SNRs) for signals in the Sangria-HM dataset.
        Waveforms are generated using FastBHB as implemented in \texttt{lisabeta} \cite{Marsat:2020rtl}.
    }
    \label{tab:characteristic_strain_snr}
\end{table}

There are 15 \ac{MBHB} signals in the Sangria-HM training dataset.
In Table~\ref{tab:characteristic_strain_snr} we summarize these signals, showing the merger
times and the signal-to-noise ratio of the full signal calculated from characteristic strain.
We also show the signal-to-noise ratios for each of the 15 \ac{MBHB} signals, at 14, 7, 4, 1 and 0.5 days before merger using frequency-domain cuts.
We see that Signal 1 is unlikely to be seen even 0.5 days pre-merger, and that only the loudest signals - 0, 3, 4, 10, 11 and possibly 12 and 13 - would be seen 4 days or more before merger.

\subsection{Smoothing the power spectral density}
\label{subsec:smoothing}
At $\sim60$\,mHz, we see a dip in the power-spectral density, this is an artifact where gravitational waves at this frequency have a wavelength
of $5\times10^{9}$\,m, which is exactly double the arm length of LISA, and the TDI combination process results in cancellation at this frequency.
Despite appearances, it is \emph{not} a point of infinite sensitivity, it is a point of zero sensitivity because any potential signal is also cancelled completely at this point.
However, a problem arises when creating a whitening filter
to ``whiten'' this because it is effectively a delta function in the inverse power-spectral density, which would require an infinitely long time-domain whitening filter to correctly whiten.

Instead, following our previous work~\cite{CabournDavies:2024hea} we choose not to whiten this feature, especially as the signals we are looking for are at frequencies two orders of magnitude
smaller than this.
To do this we simply smooth the feature out of the power spectrum.
We identify peaks in the model either side of the dip, and draw a line between them.
To avoid this process itself introducing discontinuities in the spectrum, we use a Hann window to smoothly transition between the power-spectral density
and the straight line.
We illustrate this in Figure~\ref{fig:smoothing}.

\section{Applying the Premerger Search}
\label{sec:zerolatency_sangria}

We now apply our previous zero-latency filter based search method to the Sangria-HM Training dataset to demonstrate its ability to observe premerger \acp{MBHB}.
A full description of the method can be found in~\cite{CabournDavies:2024hea} and, other than some necessary differences which we will highlight below, we use the same method, and implementation, here.

\subsection{Template Bank}
\label{subsec:template_bank}
In order to utilize a matched filter search, we must first set up a set of waveforms -- a \textit{template bank} -- to which
we compare the data.
We use stochastic bank placement~\cite{Babak:2008rb, Harry:2009ea} in the same way as~\cite{CabournDavies:2024hea},
with parameter limits set from Table~\ref{tab:parameter_space}.
\updated{These parameter choices exclude some of the massive black hole binary parameter space, particularly at lower masses.
This could be a very interesting region to explore, but is not explored in the context of proving the technique of this work.}
In constructing the bank, we calculate the match for signals only using frequencies up to the point corresponding to the desired
time-before-merger for the signal.

Although we did create such template banks in~\cite{CabournDavies:2024hea}, the inclusion of the
Galactic binary confusion noise in Sangria-HM changes the sensitivity curves and requires that we create new banks for this work. 

We create separate template banks for each of our 5 premerger search times (0.5, 1, 4, 7, and 14 days before merger) and simulate the termination of the waveforms by computing the numerical
match between waveforms only using frequencies up to frequency corresponding to the dominant gravitational-wave mode at that time-before-merger for that signal.

\begin{table}[ht]
\begin{centering}
\begin{tabular}{|c|r @{ -- } l|c|}
\hline
  Parameter & \multicolumn{2}{c|}{Limits} & Distribution\\
  \hline
  \strutrow Total Mass, $M_\odot$ & $10^6$ & $2\times 10^7$ & uniform \\
  Mass ratio, $q$ & 1 & 4 & uniform \\
  Component Spins, $\chi_1, \chi_2$ & -0.9 & 0.9 & uniform\\
  Ecliptic Longitude, $\lambda$ & 0 & $2\pi$ & uniform \\
  Ecliptic Latitude, $\beta$ & $-\pi / 2$ & $\pi / 2$ & uniform in $\sin(\beta)$\\
  Inclination, $\iota$ & 0 & $\pi / 2$ & uniform in $\cos(\iota)$ \\
  Polarisation & 0 & $2 \pi$ & uniform \\  
\hline
\end{tabular}
\caption{Parameter space used in setting up the template banks for the zero-latency filter search.
\updated{The choices made here were designed to include the full set of Sangria-HM dataset signals, which range in total mass from $1.6\times 10^6 M_\odot$ to $9.6\times 10^6 M_\odot$, mass ratio ranges from $1.03$ to $2.75$, and component spins in the z-direction range from $-0.7$ to $0.8$.}
\label{tab:parameter_space}
}
\end{centering}
\end{table}

We must define a power spectral density used for the match calculations used in deciding the stochastic placement and so in this work, we consider
the modeled power spectral density with Galactic binary confusion, and the power spectral density estimated directly from the data.
The banks used in~\cite{CabournDavies:2024hea} used a modeled power spectral density without Galactic binary confusion noise.

\begin{figure}
    \centering
    \includegraphics[width=\columnwidth]{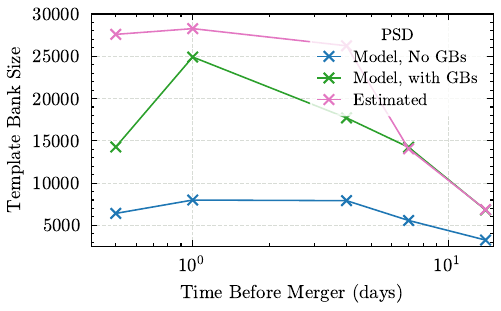}
    \caption{
        Template bank sizes assuming different power-spectral densities.
        Inclusion of unresolvable Galactic binary foreground in the model increases bank size by a factor between 2.1 and 2.6.
        Using the estimated rather than modeled power-spectral density increases bank size by a factor between 1.0 and 1.9.
    }
    \label{fig:bank_size}
\end{figure}

In Figure~\ref{fig:bank_size}, we show the bank sizes at the different time-before-merger cutoffs for the different power spectral densities.
We compute this using three of the sensitivity curves shown earlier in Figure~\ref{fig:psds}; the modeled power-spectral density with Galactic binary confusion, the modeled curve without Galactic binary confusion as used in~\cite{CabournDavies:2024hea}, and a power-spectral density estimated from data.

We can see that significantly more templates are required when the confusion noise is included, presumably due to \deleted{the signal power being accumulated over more time, rather than most of the power coming at the highest frequencies}\updated{noise at high frequency pushing the bandwidth of the template to lower frequencies.
As the longer-duration low frequency component now produces a more significant proportion of the power, small mismatches will become significant over the time, meaning more templates are needed to meet minimum match criteria}.
We also see that, for shorter times before merger, more waveforms are required for the estimated curve, than for the model curve.
In all cases the number of waveforms is relatively small, current template banks for ground based analysis are at least one order of magnitude larger than this, and would not present a computational challenge a decade from now.

For the searches in this paper, we use the template banks calculated from the estimated power-spectral density.

\subsection{Applying the search to Sangria-HM}
\label{subsec:application}

In \cite{CabournDavies:2024hea} we simulated 30 days of data, with a merger occurring at a defined point after the end of the data, and attempted to recover the signal.
Here we wish to recover the 15 signals by searching for them in the full 365 days of data.
We also wish to identify the signals as \emph{early} as possible.
To simulate this we begin by taking the first 30 days of data and searching for mergers 0.5, 1, 4, 7 and 14 days from the end of that data (ie. signals that would merge 0.5, 1, 4, 7 or 14 days in the future).
We allow a one hour search window in each case, and therefore search for signals 6 days 23 hours to 7 days premerger (and similar), but do so using waveforms that are truncated 7 days premerger.

We then read the next hour of data, and ``forget'' the first hour of data and repeat the process.
By repeating this, we walk through the entire year-long data set, hour-by-hour, simulating how data might arrive in real time while continuously searching for premerger signals.
We consider a signal to be detected if we recover it with a signal-to-noise ratio larger than 10, which is more conservative than the thresholds we computed empirically in~\cite{CabournDavies:2024hea} but allows for some uncertainty in the noise model.

\subsection{Signal removal before merger}
\label{subsec:peak_removal}

\begin{figure}
    \centering
    \includegraphics[width=\columnwidth]{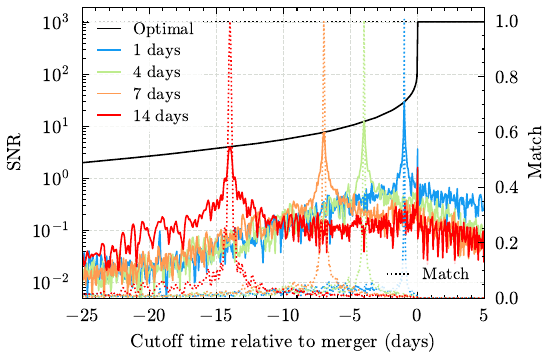}
    \caption{
        Optimal signal-to-noise ratio (SNR) build-up and signal-to-noise ratio of premerger-truncated signals matched filtered against the full signal for Signal 10.
        We see that the premerger-truncated signals' signal-to-noise ratios are a product of the optimal signal-to-noise and the match between the full and truncated signals.
        Equivalent plots for all signals are included in the data release.
    }
    \label{fig:double_peak}
\end{figure}

\begin{figure}
    \centering
    \includegraphics[width=\columnwidth]{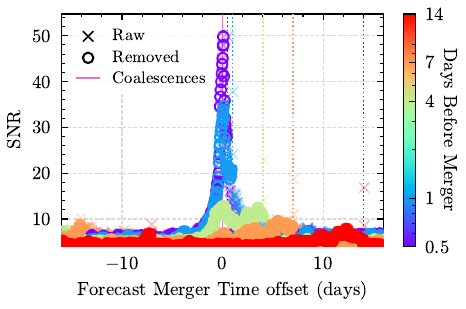}
    \caption{
        Signal 10 results, showing results with and without the secondary peaks, for the removed results,
        the signal is removed from the data two hours before merger.
        \updated{Each circle and cross correspond to a different data end time as the search is performed with the data end time stepping forward in hourly steps.}
        We also see peaks in the data before merger, these are the post-signal peaks for Signal 9, which is 21 days before Signal 10.
        Equivalent plots for all signals are included in the data release.
    }
    \label{fig:zerolatency_signal_removed_raw}
\end{figure}

Finally, before we apply the search to the Sangria-HM data we consider the cross-correlation between
the merger of a \ac{MBHB} signal and a template waveform searching for premerger waveforms.
In Figure~\ref{fig:double_peak} we show a number of things to illustrate this.
First (black solid line), we show the integrated signal-to-noise ratio of one of the signals in the Sangria-HM dataset as a function of time before (and after) merger, noting, as is well known, that the power peaks sharply in the hour leading up to merger.
Second (dotted lines) we show the normalized match between waveforms truncated 1, 4, 7 and 14 days before merger, and the signal, as a function of the signal's time before merger.
This is normalized such that a value of 1 indicates perfect agreement, and all the integrated signal-to-noise would be recovered, a value of 0 indicates complete disagreement, and no signal-to-noise ratio would be recovered, a value of 0.5 indicates that 50\% of the signal-to-noise ratio would be recovered.
This value peaks sharply to 1 at the expected times when the waveform and merger are both at the same time before merger and rapidly drops on both sides.
Finally (solid lines), we show the product of the normalized matches and the integrated signal-to-noise, representing the signal-to-noise ratio that would be recovered by searching with the waveform truncated at the stated point before merger.
We see the expected peaks at the desired time before merger.
However, we also see secondary peaks close to the merger time.
These are points where the normalized match is low, but because the integrated signal-to-noise is larger (over 100) the product of the two can result in values that are comparable, in terms of signal-to-noise, to a premerger signal.
In short, this shows that there is enough cross-correlation between pre-merger signals and loud merger events that it has the potential to confuse our search.

\begin{figure*}
    \centering
    \includegraphics[width=\textwidth]{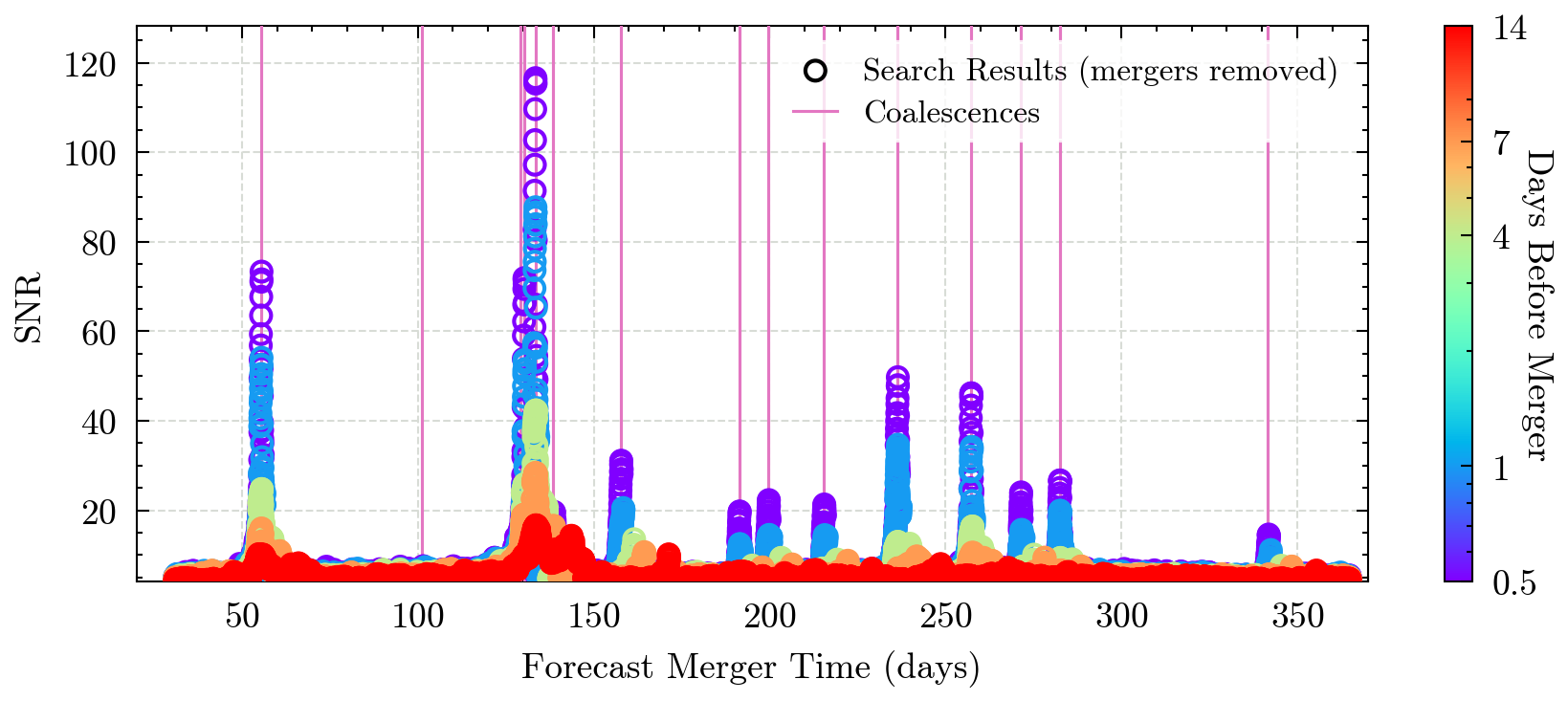}
    \caption{
        Search results for the zero-latency filter search.
        \updated{Each circle corresponds to a different data end time as the search is performed with the data end time stepping forward in hourly steps.}
    }
    \label{fig:zerolatency_remove_all}
\end{figure*}

We therefore want to avoid confusing actual mergers as premerger signals, or having a merger mask a second incoming \ac{MBHB}.
There are techniques, for example~\cite{Allen:2004gu}, that could be used to distinguish between genuine premerger signals, and premerger template waveforms overlapping with \ac{MBHB} mergers.
However, in this case, it is easier to simply remove the signals from the data before merger.
To do this, we remove each signal 2 hours before its corresponding merger time.
When removing the signal, we assume that we know the waveform perfectly and remove the full waveform from the data. 
This is a simplification; some waveforms would be identified quite a bit before 2 hours premerger, whereas some quieter signals may not be detected until close to merger, although all 15 of the signals we consider are detectable 2 hours before merger.
In a real-world scenario one could consider a scenario where we remove signals as they become detectable, which we will explore in Section~\ref{sec:focused}.
Additionally, we will not have a perfect waveform model, and there will be some level of residual due to mismatch.
However, in the real-world scenario the goal will not be perfect removal of the signal, but to remove the possibility of confusing premerger signals with loud \ac{MBHB} mergers, which would be possible with even a rough subtraction of the signal.

Figure~\ref{fig:zerolatency_signal_removed_raw} shows the effect of removing the merger when searching for Signal 10.
Without the merger removed, we ``detect'' additional mergers 14, 7, 4 and 1 day after the merger of Signal 10, due to this cross-correlation.
With the merger removed, we recover the expected result of the signal being detected first 4 days before merger with signal-to-noise increasing as the signal nears merger.

\subsection{Search Results}
\label{subsec:zerolatency_results}

Finally, we show the results from searching the full year-long dataset in Figure~\ref{fig:zerolatency_remove_all}.
We can see that peaks in the signal-to-noise ratio timeseries corresponding to all signals except Signal 1, which we did not expect to recover.
In Table~\ref{tab:when_found}, alongside our inpainting results, which we'll discuss later in Section~\ref{sec:inpainting_sangria},
we show when each signal is recovered, and with what signal-to-noise ratio. We recover all of the signals at times comparable to what was predicted in Table~\ref{tab:characteristic_strain_snr}.
We note that Signals 2 - 5 are visibly overlapping in Figure~\ref{fig:zerolatency_remove_all}.
We leave a discussion of identifying each of these as separate signals to Section~\ref{sec:focused} so that we can do so in the context of both our zero-latency and inpainting search results.

\section{Inpainting}
\label{sec:inpainting}

We have shown in the previous section, and in our previous work~\cite{CabournDavies:2024hea}, that the zero-latency filter method, used successfully for many years in ground-based low-latency analysis~\cite{Tsukada:2017cuf} is directly applicable to
solve the \ac{MBHB} premerger problem in LISA.

However, there are some issues with this method, when applied to LISA data, that warrant considering alternatives.
The premerger analysis that we carried out with the zero-latency filter here analyzes 30 days of data to identify mergers up to 14 days before merger.
LISA data is expected to have gaps arising both from planned antenna re-pointing and maneuvers as well as through unexpected outages~\cite{Colpi:2024xhw}.
Gaps lasting on the order hours are predicted to occur roughly every two weeks, with shorter, 100s long gaps expected on a daily cadence~\cite{Castelli:2024sdb}.
We do not have a method for dealing with these gaps using the zero-latency filter and therefore if there is a gap in the recorded LISA data in the days before a \ac{MBHB} merger it might result in a delay in identifying that the merger is about to happen.
Additionally, the zero-latency search we propose only runs at fixed time before merger. If a system became visible 13 days before merger our method would require waiting 6 days before a detection was made.
One could of course solve this by searching at more times than we do here, but that comes at an increase in computational cost.

Here we consider an alternative technique, which is also adapted from ground-based analyses.
In particular the observation of the first binary neutron star merger, GW170817, was significantly hampered by the presence of a loud non-Gaussian artifact in the seconds before the merger.
At the time of observation the glitch was manually excised from the data, in a process known as ``gating".
While this gating removes the immediate artifact, it introduces spectral leakage in the whitened strain, as described in~\cite{Zackay:2019kkv}, which can cause issues for detection pipelines and bias inference algorithms.

A method for first removing and then ``inpainting'' these non-Gaussianities was proposed in~\cite{Zackay:2019kkv} and can equivalently be directly applied for analyzing over data gaps.
The full derivation is given in~\cite{Zackay:2019kkv,Capano:2021etf} and we refer the interested reader there for a full description, here we briefly cover the main principle.
Consider the matched-filter between signal, $s$ and data $h$
\begin{equation}
    (h|s) = 4\mathrm{Re}\int_{f_\mathrm{low}}^{f_\mathrm{high}}\frac{\tilde{h}(f)\tilde{s}^*(f)}{S_n(f)} \mathrm{d} f,
\end{equation}
we can rewrite this as
\begin{equation}
    (h|s) = 4\mathrm{Re}\int_{f_\mathrm{low}}^{f_\mathrm{high}} \tilde{h}(f) \tilde{u}(f)  \mathrm{d} f,
\end{equation}
where $\tilde{u}(f) = \tilde{s}^*(f) / S_n(f)$ is the ``overwhitened'' strain data.
The matched-filter is then simply the cross correlation between the overwhitened strain and the template waveform and so any times where the overwhitened strain is 0 do not contribute to the overall matched-filter.
However, if we begin with raw, unwhitened, data with a gap in it, the whitening, or overwhitening, process will ``bleed'' data from the valid points into the gap according to the length of the [over]whitening filter in the time-domain.
This results in the values of the overwhitened data in the gap not being zero.
The inpainting technique works by choosing appropriate values of the unwhitened data, in the gap, such that after overwhitening all values in the gap \emph{are} zero.
Choosing values that the unwhitened data should take in the gap to achieve this boils down to a linear algebra problem, which can be solved by a Toeplitz solver~\cite{Zackay:2019kkv}.

\subsection{Implementing inpainting to simultaneously search for a signal at all times}

The ``inpainting'' procedure is implemented in the \texttt{PyCBC} toolkit as described in~\cite{Capano:2021etf}.
We first describe how we use this to inpaint the raw strain, and then how we perform matched-filtering on the inpainted data to obtain signal-to-noise ratio as a function of the time before merger.

To apply \texttt{PyCBC}'s inpainting to our search, where we have gaps in the middle of the data we simply apply it directly, allowing it to select values in the raw strain to make the overwhitened strain in the gap 0.
To be able to analyze data that ends abruptly, where we are searching for signals merging in the future, we add a long stretch of zeroes to the end of the data, adding at least a week of extra zeroed data.
We then apply the inpainting over this ``gap'', simultaneously dealing with any issues due to the abrupt end of the data, but also smoothing out the abrupt start.

With the inpainting approach we can simultaneously search for a signal matching a given template waveform at all times before merger.
It is common in ground based detection analyses~\cite{Allen:2005fk} to evaluate $(h|s)$ as a function of time by introducing a frequency dependent time shift and evaluating the resulting expression using a Fast Fourier Transform.
\begin{equation}
    (h|s)(t) = 4\mathrm{Re}\int_{f_\mathrm{low}}^{f_\mathrm{high}} \tilde{h}(f) \frac{\tilde{s}^*(f)}{S_n(f)} e^{2 i \pi f t} \mathrm{d} f.
\end{equation}
This can be directly applied here as well. Where the signal waveform overlaps with the zeroed-out
overwhitened data there is no contribution to $(h|s)$ and so we can easily obtain $(h|s)$ as a function of shifting merger time.
However, there is a complication. For this to be interpreted as a signal-to-noise ratio we need to normalize by the template amplitude; $\rho = (h|s) / (h|h)^{0.5}$.
In this application, the template amplitude $(h|h)$ must now be modeled as time dependent because any power in the template waveform at times where there is no data should not be included.
In~\cite{Zackay:2019kkv} a method for computing a time-dependent $(h|h)$ was proposed, and we adopt it here.
Consider computing $(h|h)$ in the time domain, where $(h|h)$, without gaps, is computed as the integral of the whitened template waveform multiplied by itself.
Then $(h|h)$ can be computed at any given value of the merger time by zeroing out the parts of the template waveform that lie in the gaps.
To obtain $(h|h)(t)$ one can compute the cross-correlation between a vector whose values are 1 where data is present and 0 where there are gaps, and the whitened template waveform multiplied by itself.
This is more efficiently computed in the frequency domain via the convolution theorem
\begin{equation}
    (h|h)(t) = 4\mathrm{Re}\int \tilde{h_2}(f) \tilde{z}(f) e^{2 i \pi f t} \mathrm{d} f,
\end{equation}
where $\tilde{h_2}(f)$ is the Fourier transform of the whitened template waveform multiplied by itself and $\tilde{z}$ is the Fourier transform
of the vector containing 1 where data is present and 0 where there are gaps.
This gives us everything we need to compute $\rho(t)$ given a template waveform that might merge in the future and with the presence of data gaps.

\begin{figure*}
    \centering
    \includegraphics[width=\textwidth]{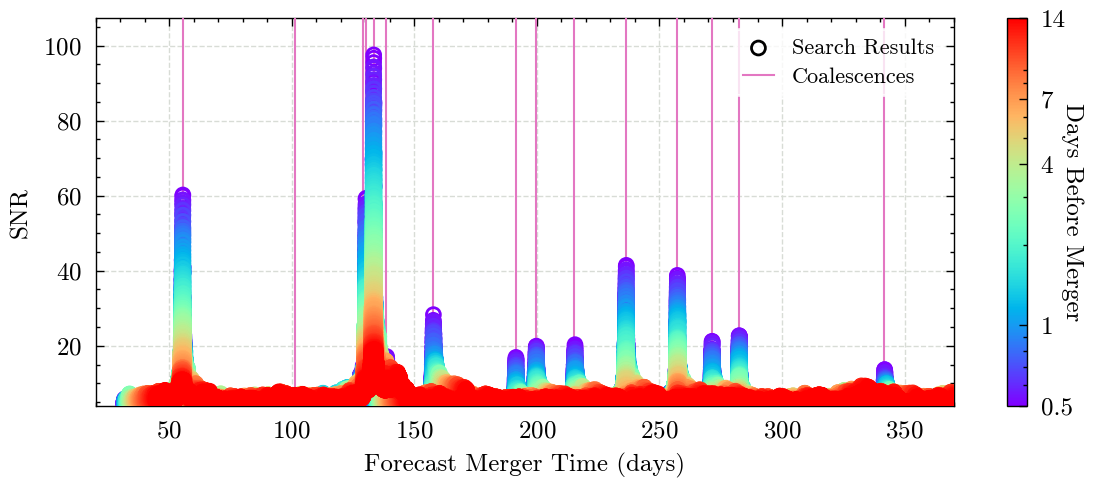}
    \caption{
        Signal-to-noise ratio (SNR) as a function of the forecast merger time and the time before merger. 
        \updated{Each circle corresponds to a different data end time as the search is performed with the data end time stepping forward in hourly steps.}
        Produced using the inpainting technique to analyze the Sangria-HM dataset.
        Signals have been removed from the data once they reach 2 hours before merger.
    }
    \label{fig:inpainting_estimate}
\end{figure*}

\subsection{A template bank to search for a signal at all times}

The final step to be able to apply inpainting to a premerger search is that we need to create a set of filter waveforms, or template bank.
The template banks created for the zero-latency search are created assuming a signal either 0.5, 1, 4, 7 or 14 days before merger and so are not guaranteed to be applicable here.
To demonstrate this method here we simply use the template bank computed assuming a signal 1 day before merger, the largest of the five banks produced.

An alternative method, which we encourage to be explored for future applications of this method, would be to create a template bank that is effectual for observing any signals in a given range of times before merger.
Stochastic placement, as we used for the zero-latency search~\cite{CabournDavies:2024hea}, could be used to build such a bank.
However, when simulating signals we would draw the time before merger when selecting a proposal point, alongside the other parameters.
We would then check if that signal would be recovered by the existing template bank and add it to the bank if it would not be recovered below a specified match threshold.

\section{Applying inpainting to search Sangria-HM data}
\label{sec:inpainting_sangria}

\begin{table*}[tbp]
        \centering
\begin{tabular}{|c|cc|cc|c|}
\hline
\multirow{3}{*}{Signal Number} & \multicolumn{4}{c|}{Found} & \multirow{3}{*}{Extra Warning (days)} \\
\cline{2-5}
& \multicolumn{2}{c|}{Zero Latency} & \multicolumn{2}{c|}{Inpainting} & \\
\cline{2-5}
 & Time (days)  & SNR & Time (days)  & SNR & \\
\hline
0 & 7 &   14.665 & 14.00 &  10.632 & 7.00 \\
1 & \multicolumn{2}{c|}{Never}    & \multicolumn{2}{c|}{Never}   & N/A \\
6 & 1 &   14.343 & 1.67 &  10.191 & 0.67 \\
7 & 0.5 &   15.838 & 0.96 &  10.346 & 0.46 \\
8 & 1 &   12.596 & 1.50 &  10.151 & 0.50 \\
9 & 1 &   12.051 & 1.33 &  10.101 & 0.33 \\
10 & 4 &   11.870 & 4.96 &  10.188 & 0.96 \\
11 & 4 &   13.896 & 6.71 &  10.071 & 2.71 \\
12 & 1 &   12.188 & 1.58 &  10.241 & 0.58 \\
13 & 1 &   17.453 & 2.67 &  10.118 & 1.67 \\
14 & 0.5 &   11.177 & 0.54 &  10.590 & 0.04 \\
\hline
\end{tabular}
        \caption{
            Results from the search on Sangria-HM, indicating when signals first cross the signal-to-noise ratio threshold of 10.
            Rows for Signals 2--5 are not shown as the signals overlap.
            We discuss further how we disentangle these signals in Section~\ref{sec:focused}.
        }
        \label{tab:when_found}
\end{table*}

Having established the methodology by which we will use the inpainting technique to search for premerger signals, we apply the method to the Sangria-HM dataset.
We use the same dataset as we used earlier for the zero-latency filter and search for signals that are between 14 days and 0.5 days before merger.
In Section~\ref{sec:zerolatency_sangria}, we remove signals 2 hours before merger so that a loud merger does not interfere with rapid observation of subsequent signals.

The results are presented in Figure~\ref{fig:inpainting_estimate} where we show signal-to-noise ratio as a function of the time before merger.
We also show in Table~\ref{tab:when_found} the details of how long before merger each signal was identified and with what signal-to-noise ratio.
This is compared to the previous zero-latency results.
The main thing we observe from this is that the inpainting procedure is doing what we expect, recovering signals at roughly the time before merger that we expect from Table~\ref{tab:characteristic_strain_snr}.
We expect some variation due to discreteness of the template banks used, noise realization and that the signal-to-noise ratios given in Table~\ref{tab:characteristic_strain_snr} are produced with a frequency-domain cut, rather than an abrupt termination in the time domain. 

We notice that as the inpainting search is not restricted to search at \deleted{discrete}\updated{a small set of} times before merger we are able to detect some of the signals days earlier than our zero-latency search.
The zero-latency search \emph{could} be run at more times before merger to alleviate this, but the computational cost would grow linearly with the number of times before merger searched.

We also observe that for Signals 0 and 3 the zero-latency search identifies the signal 7 days before merger, whereas the inpainting search identifies it 14 days before merger.
Additionally, Signal 4 is found with a larger signal-to-noise ratio with the inpainting search 14 days before merger.
We believe that this improved efficiency at 14 days before merger is due to how the data is conditioned.
For the zero-latency search we apply a long taper to the waveform over the first $\sim$ day of data to avoid any abrupt start effects and to remove any wraparound effects from the ringdown.
The same is not performed for the inpainting. This results in more power being available for inpainting search when searching 14 days before merger.
This is not a fundamental limitation; we chose to demonstrate the zero-latency method, and inpainting, on 28 days of data, but if more data was available, with non-negligible power there, one could simply increase the analysis window.
However, for the zero-latency search, this increases the chance of gaps being present, which it cannot handle.

\subsection{Gaps}
\label{subsec:inpainting_gaps}

\begin{figure}
    \centering
    \includegraphics[width=\columnwidth]{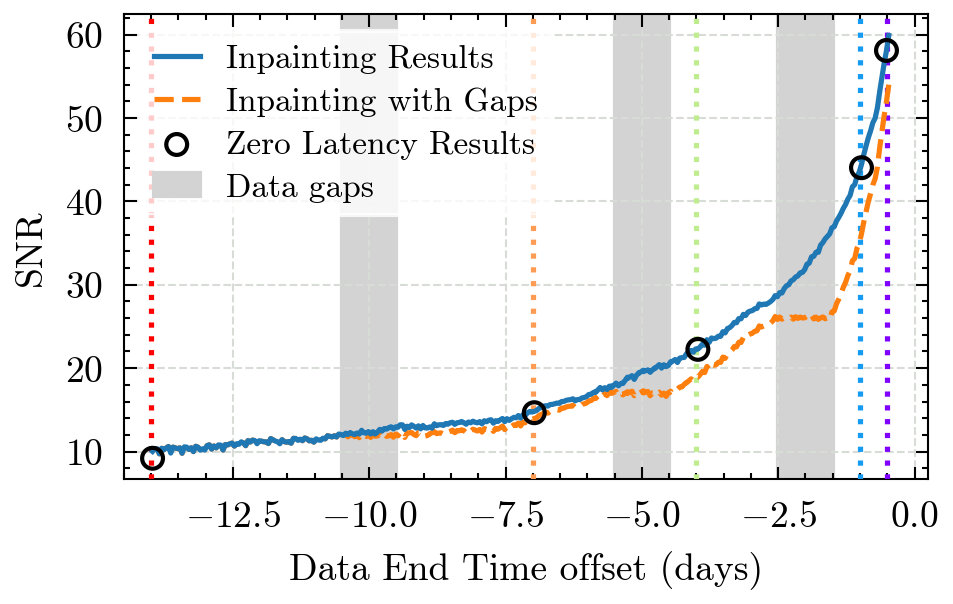}
    \caption{
        Timeline of results for Signal Zero, demonstrating the inclusion of gaps in the data in the days before merger.
        Timeline plots for all signals are included in the data release.
    }
    \label{fig:signal_zero_timeline}
\end{figure}

One of the main motivations for the inpainting approach was that it could effectively handle gaps in the data.
To demonstrate this, we reanalyse Signal 0 with gaps added to the data 10.5 - 9.5, 5.5 - 4.5 and 2.5 - 1.5 days before merger.
We then search for Signal 0 using inpainting to mitigate the gaps and show the results in Figure~\ref{fig:signal_zero_timeline}.

The plot shows the result of searching for the signal with the zero-latency search without gaps, the inpainting search without gaps and the inpainting search with the aforementioned gaps added.
We note that the overall signal-to-noise ratio recovered when gaps are present is smaller--which is expected, because data, and therefore signal power, is missing--but the search still accrues signal-to-noise at times when data is available, maximizing the chance of finding a signal like this.
If, for example, Signal 0 was four times further away, such that the signal-to-noise was reduced by a factor of four compared to what we see, the inpainting or zero-latency search could find this signal 1.22 days before merger.
However, if the data contains gaps then the zero-latency search would not be able to identify this 
signal at all with our current implementation, whereas the inpainting search would still find it 0.81 days before merger.

\section{Overlapping Signals}
\label{sec:focused}

\begin{figure*}
    \centering
    \begin{subfigure}{\columnwidth}
    \includegraphics[width=\columnwidth]{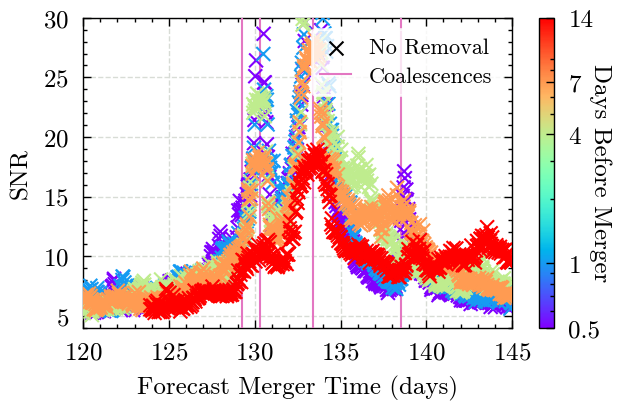}
    \caption{No removal of signals}
    \label{fig:no_removal}
    \end{subfigure}
    \begin{subfigure}{\columnwidth}
    \includegraphics[width=\columnwidth]{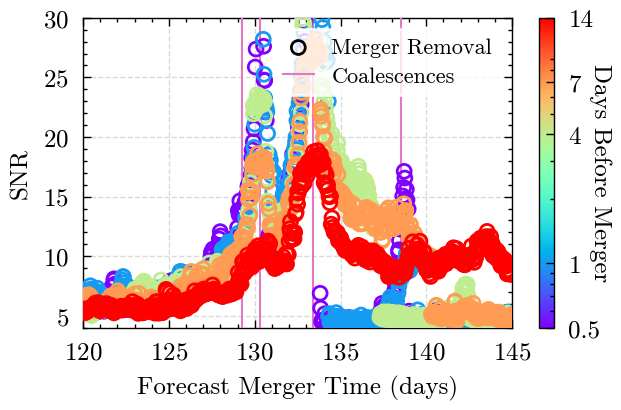}
    \caption{Signals are removed 2 hours before merger}
    \label{fig:2hr_removal}
    \end{subfigure}
    \caption{
        Signal-to-Noise Ratio (SNR) vs forecast end time results for the inpainting search in the congested region of days 120-145.
        \updated{Each circle and cross correspond to a different data end time as the search is performed with the data end time stepping forward in hourly steps.}
        For the left plot (\ref{fig:no_removal}), we do not remove the signal from the data at any point.
        We see overlap between signals, where, for example, Signal 2 is not visible as the ramp-up in signal-to-noise ratio for Signal 3 is dominant.
        The Signal 5 results are also overwhelmed by the ramp-down of Signal 4.
        For the right plot (\ref{fig:2hr_removal}), we remove the signal 2 hours before merger, and we see that the situation is not visibly improved.
        We use an upper limit on the signal-to-noise ratio axis of 30 so that we can see features clearly - the signal-to-noise ratio increases to around one hundred for the 0.5 days before merger case (See Figure~\ref{fig:inpainting_estimate}).
    }
    \label{fig:focused_no_2hr_removal}
\end{figure*}

The Sangria-HM training dataset that we use contains 15 MBHB signals with end times chosen uniformly within the dataset.
This resulted in 4 signals, Signals 2 - 5, having end times that are all within 10 days of each other, with Signals 2 and 3 separated by only 1.05 days.
This proximity of signals can lead to some overlap between them.
In Figure~\ref{fig:focused_no_2hr_removal} we show the results, without any signal removal (\ref{fig:no_removal}), or with signal removal 2 hours before merger (\ref{fig:2hr_removal}) for the inpainting search.

We can clearly see a lot of cross talk between the signals, as Signals 3 and 4 cause long ramps up and down in signal-to-noise ratio before and after their signal peak.
In Figure~\ref{fig:zerolatency_signal_removed_raw} we demonstrated the need to remove signals before merger to help avoid post-signal residual power, which is seen in Figure~\ref{fig:2hr_removal}.
The distinction between signals is clearer now; Signal 5 can now be seen in the one-day-before-merger results.
However it is still insufficient to distinguish Signals 3 from 4 and both Signals 2 and 5 remain masked by the louder signals.

This highlights that it is important not just to remove the merger of these MBHB signals from the data when searching for other MBHBs, but to remove loud signals from the data as they are observed to be able to reliably identify multiple signals in the data.
To remove such signals from the data will require low-latency premerger inference of the properties of the signals, which is not the scope of this paper.
Here we assume that we are able to reliably model the signals found in the data and search again over the region containing Signals 2 - 5, using the inpainting search and removing signals at the times defined in Table~\ref{tab:focused_removal_times}.

\begin{table}
\begin{tabular}{|cc|cc|}
\hline
\multicolumn{2}{|c|}{Signal} & \multicolumn{2}{c|}{Removal Time} \\
\hline
Number & Time (days) & Offset & Time (days) \\
\hline
2 & 129.3 & 2 hours & 129.2 \\
3 & 130.3 & 11 days & 119.3 \\
4 & 133.4 & 14 days & 119.4 \\
5 & 138.6 & 2 hours & 138.5 \\
\hline
\end{tabular}
\caption{
    Times-before-merger at which we remove signals from the data in the focused region.
    Signal 4 is removed 14 days before merger, which corresponds to 10.9 days before Signal 3's merger.
    Signals 3 and 4 are removed 9.95 and 9.84 days before Signal 2 merges respectively.
    \label{tab:focused_removal_times}
}
\end{table}

These times are a little more conservative than those quoted in Table~\ref{tab:when_found} which indicates where the signal-to-noise ratio goes above ten.
This reflects human uncertainty of initially identifying Signals 3 and 4 as separate signals and similarly disentangling Signals 2 and 5 from the louder signals between them.

\begin{figure*}
    \centering
    \includegraphics[width=1.5\columnwidth]{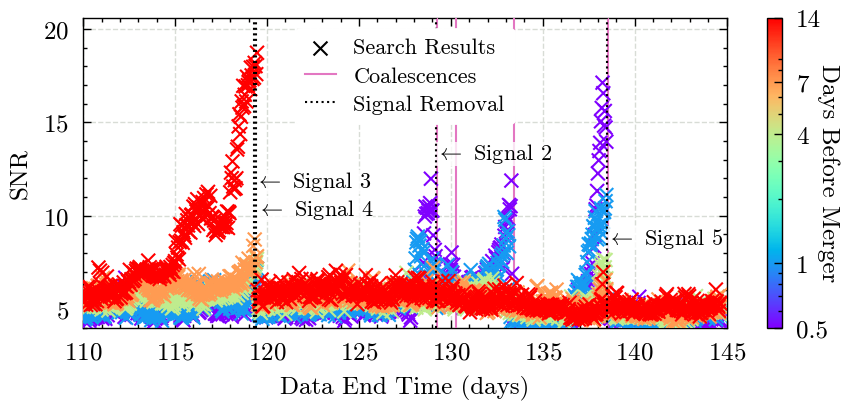}
    \caption{
        Timeline of results in the congested period of days 110-140 of the Sangria-HM dataset.
        \updated{Each and cross corresponds to a different data end time as the search is performed with the data end time stepping forward in hourly steps.}
        These are inpainting results filtered to be at the 0, 5, 1, 4, 7 and 14 days-before-merger that were used in the zero latency search.
        We see that Signal 4 - the loudest signal - dominates the SNR, and can actually be found before Signals 2 and 3.
        We remove Signal 4 14 days before merger.
        Signal 3 can be detected almost immediately after Signal 4 is removed from the data, and is therefore removed 11 days before merger.
        Signal 2 can then be found around a day before merger.
        Peaks remain for the 1 and 0.5 days-before-merger results for Signals 3 and 4 - this is due to \deleted{imperfect removal of the signal from the data}\updated{a bug in the signal removal, and is explained fully in the data release}.
    }
    \label{fig:variable_removal}
\end{figure*}

We then show results of this search, removing signals as they are identified, in Figure~\ref{fig:variable_removal}.

This is now showing the signal identification as a function of wall clock time, as opposed to the predicted merger time used in other plots.
This is done to highlight the asynchronous order in which the signals are identified.
In particular, once we reach Signal 4 being 14 days out of merger it is possible to clearly identify that two separate signals (3 and 4) are present in the data, which are then removed at almost the same point.
Following this, almost 10 days later, Signal 2 is identified as it nears merger, before Signal 5 is observed also within a day of it merging.
We see here that more realistic results for when Signals 2, 3, 4 and 5 cross the SNR threshold of 10 would be 0.5, 7, 14 and 0.5 respectively.
For Signal 3 this is significantly after the estimated time in Table~\ref{tab:characteristic_strain_snr} because of cross-talk with Signal 4.

\updated{An additional peak is seen at the time when Signal 4 merges, this is erroneous, although we are removing Signal 4 using identical parameters to how it was added to the data we did not reproduce the waveform with complete accuracy, due to a bug which leaves a noticeable residual at the merger of this loud signal.
This is explored further in the data release.}

\section{Conclusion and Discussion}
\label{sec:conclusion}

We have demonstrated the application of two separate search techniques, one using a zero-latency
whitening filter, and the other using inpainting, to identify premerger \ac{MBHB} signals in the Sangria-HM Training dataset.
Both methods recovered signals at the times predicted by noiseless calculations of the accumulation of signal-to-noise ratio as these signals approach merger.
While the Sangria-HM dataset does not include gaps, we demonstrate that our inpainting technique is capable of recovering premerger signals in the presence of gaps, which are expected to exist in the 30 day period of real LISA data that we use for our search.

We believe that the methods presented here, being of negligible computing cost for the late 2030s and utilizing similar signal processing techniques to those developed over the last decades for ground-based analyses represent viable search strategies for premerger observation of \ac{MBHB} after LISA launches.
The authors are keen to continue to work closely with the LISA Ground Segment, and the ``L2A'' package in particular, to apply these methods to more realistic simulations of LISA data that the LISA Simulation Group will produce in the coming years and tackle any further difficulties that analysis of more realistic data (e.g. correlated noise, non-Gaussianities, or data arriving in a non-uniform manner) will require.

\begin{acknowledgments}
For the purpose of open access, the authors have applied a Creative Commons Attribution (CC-BY)
licence to any author accepted manuscript version arising from this submission.

GCD and IH acknowledge support from the UKSA, through grant UKRI968,
supporting the UK's contribution to LISA's Ground
Segment activities.

The authors are grateful for computational resources provided
by Cardiff University, and funded by STFC awards supporting
UK Involvement in the Operation of Advanced LIGO.
Numerical computations were also carried out on the Sciama High
Performance Computing (HPC) cluster, which is supported by the
Institute of Cosmology and Gravitation, the South-East Physics
Network (SEPNet) and the University of Portsmouth.

\textit{Software} This work made use of the following software packages:
\texttt{BBHx}~\cite{michael_katz_2023_7791640},
\texttt{GstLAL}~\cite{Cannon:2020qnf},
\texttt{lisabeta} \cite{Marsat:2020rtl},
\texttt{matplotlib}~\cite{Hunter:2007},
\texttt{numpy}~\cite{harris2020array},
\texttt{PyCBC}~\cite{alex_nitz_2024_10473621},
\texttt{scipy}~\cite{2020SciPy-NMeth}.

Large Language Models were used in this work to generate code snippets efficiently,
debug errors, and provide editing assistance in the writing of this manuscript.

\end{acknowledgments}

\appendix

\section{Power loss due to resolvable Galactic Binary foreground}
\label{sec:gb_noise_snr_loss}
\updated{As demonstrated in Figure~\ref{fig:psds}, the pre-merger analysis waveforms mainly terminate before the frequency where resolvable galactic binary noise contribution becomes significant.
To investigate this, we compare the expected signal-to-noise ratio loss when using different PSDs.
Firstly, we use the estimated Sangria-HM `day zero' PSD, which includes signals from all sources, this is `A Estimated' in Figure~\ref{fig:psds}.
We compare to the modelled PSD, `Model A/E', which contains unresolvable galactic binary foreground which would not be removed by any global fit analyses.}

\begin{figure}
   \centering
   \includegraphics[width=\columnwidth]{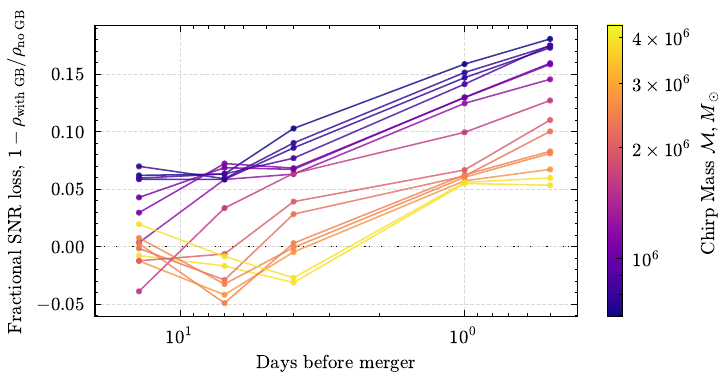}
   \caption{
       \updated{Fractional signal-to-noise ratio differences for each signal when using the `A Estimated' and `Model A/E' PSDs from Figure~\ref{fig:psds}.}
   }
   \label{fig:gb_snr_loss}
\end{figure}

\updated{The fraction of signal-to-noise ratio lost is shown in Figure~\ref{fig:gb_snr_loss}.
We see that for the signals at lower chirp masses, we can lose up to 17\% of power half a day before merger.
For heavier signals, and for times more than a few days before merger, the signal power loss is much less.
The few points where there seems to be a power `gain' (loss $<$ 0) are due to the inherent variability in estimating a PSD.
The signal-to-noise ratio loss is enough to show that global fit-style subtraction of resolved Galactic Binary sources will help optimise this technique, particularly as we move closer to merger, but is not significant enough that this subtraction is strictly necessary for pre-merger searches.
This finding is similar to what is shown in Section~\ref{sec:focused}, that the removal of signals from the data is desirable in some cases.
}

\bibliography{bibliography}

@misc{LISA:2017pwj,
    author = "Amaro-Seoane, Pau and others",
    collaboration = "LISA",
    title = "{Laser Interferometer Space Antenna}",
    eprint = "1702.00786",
    archivePrefix = "arXiv",
    primaryClass = "astro-ph.IM",
    month = "2",
    year = "2017"
}

@article{LISA:2022yao,
    author = "Seoane, Pau Amaro and others",
    collaboration = "LISA",
    title = "{Astrophysics with the Laser Interferometer Space Antenna}",
    eprint = "2203.06016",
    archivePrefix = "arXiv",
    primaryClass = "gr-qc",
    doi = "10.1007/s41114-022-00041-y",
    journal = "Living Rev. Rel.",
    volume = "26",
    number = "1",
    pages = "2",
    year = "2023"
}

@article{Arnaud:2006gm,
    author = "Arnaud, Keith A. and others",
    editor = "Merkowitz, Stephen M. and Livas, Jeffrey C.",
    title = "{The Mock LISA Data Challenges: An overview}",
    eprint = "gr-qc/0609105",
    archivePrefix = "arXiv",
    doi = "10.1063/1.2405108",
    journal = "AIP Conf. Proc.",
    volume = "873",
    number = "1",
    pages = "619--624",
    year = "2006"
}

@article{MockLISADataChallengeTaskForce:2006sgi,
    author = "Arnaud, Keith A. and others",
    editor = "Merkowitz, Stephen M. and Livas, Jeffrey C.",
    collaboration = "Mock LISA Data Challenge Task Force",
    title = "{A How-To for the Mock LISA Data Challenges}",
    eprint = "gr-qc/0609106",
    archivePrefix = "arXiv",
    doi = "10.1063/1.2405109",
    journal = "AIP Conf. Proc.",
    volume = "873",
    number = "1",
    pages = "625--632",
    year = "2006"
}

@article{Arnaud:2007vr,
    author = "Arnaud, K. A. and others",
    editor = "Krishnan, B. and Papa, M. A. and Schutz, Bernard F.",
    title = "{Report on the first round of the Mock LISA data challenges}",
    eprint = "gr-qc/0701139",
    archivePrefix = "arXiv",
    doi = "10.1088/0264-9381/24/19/S16",
    journal = "Class. Quant. Grav.",
    volume = "24",
    pages = "S529--S540",
    year = "2007"
}

@article{MockLISADataChallengeTaskForce:2007iof,
    author = "Babak, Stanislav and others",
    editor = "Scott, Susan M. and McClelland, David E.",
    collaboration = "Mock LISA Data Challenge Task Force",
    title = "{Report on the second Mock LISA Data Challenge}",
    eprint = "0711.2667",
    archivePrefix = "arXiv",
    primaryClass = "gr-qc",
    doi = "10.1088/0264-9381/25/11/114037",
    journal = "Class. Quant. Grav.",
    volume = "25",
    pages = "114037",
    year = "2008"
}

@article{Babak:2008aa,
    author = "Babak, Stanislav and others",
    editor = "Hughes, S. and Katsavounidis, E.",
    title = "{The Mock LISA Data Challenges: From Challenge 1B to Challenge 3}",
    eprint = "0806.2110",
    archivePrefix = "arXiv",
    primaryClass = "gr-qc",
    doi = "10.1088/0264-9381/25/18/184026",
    journal = "Class. Quant. Grav.",
    volume = "25",
    pages = "184026",
    year = "2008"
}

@article{Arnaud:2007jy,
    author = "Arnaud, K. A. and others",
    editor = "Krishnan, B. and Papa, M. A. and Schutz, Bernard F.",
    title = "{An Overview of the second round of the Mock LISA Data Challenges}",
    eprint = "gr-qc/0701170",
    archivePrefix = "arXiv",
    doi = "10.1088/0264-9381/24/19/S18",
    journal = "Class. Quant. Grav.",
    volume = "24",
    pages = "S551--S564",
    year = "2007"
}

@article{MockLISADataChallengeTaskForce:2009wir,
    author = "Babak, Stanislav and others",
    editor = "Marka, Zsuzsa and Marka, Szabolcs",
    collaboration = "Mock LISA Data Challenge Task Force",
    title = "{The Mock LISA Data Challenges: From Challenge 3 to Challenge 4}",
    eprint = "0912.0548",
    archivePrefix = "arXiv",
    primaryClass = "gr-qc",
    doi = "10.1088/0264-9381/27/8/084009",
    journal = "Class. Quant. Grav.",
    volume = "27",
    pages = "084009",
    year = "2010"
}

@misc{LDC_WEBSITE,
  title = "{The New LISA Data Challenges}",
  howpublished = {\url{https://lisa-ldc.lal.in2p3.fr/}},
}

@misc{LDC_WEBSITE_SANGRIA_HM,
  title = "{The New LISA Data Challenges}",
  howpublished = {\url{https://lisa-ldc.in2p3.fr/challenge2a}},
}

@inproceedings{Baghi:2022ucj,
    author = "Baghi, Quentin",
    collaboration = "LDC Working Group",
    title = "{The LISA Data Challenges}",
    booktitle = "{56th Rencontres de Moriond on Gravitation}",
    eprint = "2204.12142",
    archivePrefix = "arXiv",
    primaryClass = "gr-qc",
    month = "4",
    year = "2022"
}

@article{Allen:2005fk,
    author = "Allen, Bruce and Anderson, Warren G. and Brady, Patrick R. and Brown, Duncan A. and Creighton, Jolien D. E.",
    title = "{FINDCHIRP: An Algorithm for detection of gravitational waves from inspiraling compact binaries}",
    eprint = "gr-qc/0509116",
    archivePrefix = "arXiv",
    doi = "10.1103/PhysRevD.85.122006",
    journal = "Phys. Rev. D",
    volume = "85",
    pages = "122006",
    year = "2012"
}

@article{Babak:2008rb,
    author = "Babak, Stanislav",
    title = "{Building a stochastic template bank for detecting massive black hole binaries}",
    eprint = "0801.4070",
    archivePrefix = "arXiv",
    primaryClass = "gr-qc",
    doi = "10.1088/0264-9381/25/19/195011",
    journal = "Class. Quant. Grav.",
    volume = "25",
    pages = "195011",
    year = "2008"
}

@article{CabournDavies:2024hea,
    author = "{Cabourn Davies}, Gareth and others",
    title = "{Premerger observation and characterization of massive black hole binaries}",
    eprint = "2411.07020",
    archivePrefix = "arXiv",
    primaryClass = "hep-ex",
    doi = "10.1103/PhysRevD.111.043045",
    journal = "Phys. Rev. D",
    volume = "111",
    number = "4",
    pages = "043045",
    year = "2025"
}

@article{Cannon:2020qnf,
title = {GstLAL: A software framework for gravitational wave discovery},
journal = {SoftwareX},
volume = {14},
pages = {100680},
year = {2021},
issn = {2352-7110},
doi = {https://doi.org/10.1016/j.softx.2021.100680},
url = {https://www.sciencedirect.com/science/article/pii/S235271102100025X},
author = {Kipp Cannon and Sarah Caudill and Chiwai Chan and Bryce Cousins and Jolien D.E. Creighton and Becca Ewing and Heather Fong and Patrick Godwin and Chad Hanna and Shaun Hooper and Rachael Huxford and Ryan Magee and Duncan Meacher and Cody Messick and Soichiro Morisaki and Debnandini Mukherjee and Hiroaki Ohta and Alexander Pace and Stephen Privitera and Iris {de Ruiter} and Surabhi Sachdev and Leo Singer and Divya Singh and Ron Tapia and Leo Tsukada and Daichi Tsuna and Takuya Tsutsui and Koh Ueno and Aaron Viets and Leslie Wade and Madeline Wade},
eprint = "2010.05082",
archivePrefix = "arXiv",
primaryClass = "astro-ph.IM"
}

@article{Garcia-Quiros:2020qpx,
    author = "Garc{\'\i}a-Quir{\'o}s, Cecilio and Colleoni, Marta and Husa, Sascha and Estell{\'e}s, H{\'e}ctor and Pratten, Geraint and Ramos-Buades, Antoni and Mateu-Lucena, Maite and Jaume, Rafel",
    title = "{Multimode frequency-domain model for the gravitational wave signal from nonprecessing black-hole binaries}",
    eprint = "2001.10914",
    archivePrefix = "arXiv",
    primaryClass = "gr-qc",
    doi = "10.1103/PhysRevD.102.064002",
    journal = "Phys. Rev. D",
    volume = "102",
    number = "6",
    pages = "064002",
    year = "2020"
}

@article{Harry:2009ea,
    author = "Harry, Ian W. and Allen, Bruce and Sathyaprakash, B. S.",
    title = "{A Stochastic template placement algorithm for gravitational wave data analysis}",
    eprint = "0908.2090",
    archivePrefix = "arXiv",
    primaryClass = "gr-qc",
    doi = "10.1103/PhysRevD.80.104014",
    journal = "Phys. Rev. D",
    volume = "80",
    pages = "104014",
    year = "2009"
}

@misc{michael_katz_2023_7791640,
  author       = {Michael Katz and
                  Jonathan Roberts},
  title        = {mikekatz04/BBHx: New release!},
  month        = apr,
  year         = 2023,
  publisher    = {Zenodo},
  version      = {v1.0.5},
  doi          = {10.5281/zenodo.7791640},
  url          = {https://doi.org/10.5281/zenodo.7791640}
}

@article{Moore:2014lga,
    author = "Moore, C. J. and Cole, R. H. and Berry, C. P. L.",
    title = "{Gravitational-wave sensitivity curves}",
    eprint = "1408.0740",
    archivePrefix = "arXiv",
    primaryClass = "gr-qc",
    reportNumber = "LIGO-P1400129",
    doi = "10.1088/0264-9381/32/1/015014",
    journal = "Class. Quant. Grav.",
    volume = "32",
    number = "1",
    pages = "015014",
    year = "2015"
}

@article{Tsukada:2017cuf,
    author = "Tsukada, Leo and Cannon, Kipp and Hanna, Chad and Keppel, Drew and Meacher, Duncan and Messick, Cody",
    title = "{Application of a Zero-latency Whitening Filter to Compact Binary Coalescence Gravitational-wave Searches}",
    eprint = "1708.04125",
    archivePrefix = "arXiv",
    primaryClass = "astro-ph.IM",
    doi = "10.1103/PhysRevD.97.103009",
    journal = "Phys. Rev. D",
    volume = "97",
    number = "10",
    pages = "103009",
    year = "2018"
}

@article{Saini:2022hrs,
    author = "Saini, Pankaj and Bhat, Sajad A. and Arun, K. G.",
    title = "{Premerger localization of intermediate mass binary black holes with LISA and prospects of joint observations with Athena and LSST}",
    eprint = "2208.03004",
    archivePrefix = "arXiv",
    primaryClass = "gr-qc",
    doi = "10.1103/PhysRevD.106.104015",
    journal = "Phys. Rev. D",
    volume = "106",
    number = "10",
    pages = "104015",
    year = "2022"
}

@article{Houba:2024mqj,
    author = "Houba, Niklas and Strub, Stefan H. and Ferraioli, Luigi and Giardini, Domenico",
    title = "{Detection and prediction of future massive black hole mergers with machine learning and truncated waveforms}",
    eprint = "2405.11340",
    archivePrefix = "arXiv",
    primaryClass = "astro-ph.IM",
    doi = "10.1103/PhysRevD.110.062003",
    journal = "Phys. Rev. D",
    volume = "110",
    number = "6",
    pages = "062003",
    year = "2024"
}

@article{Kocsis:2007yu,
    author = "Kocsis, Bence and Haiman, Zoltan and Menou, Kristen",
    title = "{Pre-Merger Localization of Gravitational-Wave Standard Sirens With LISA: Triggered Search for an Electromagnetic Counterpart}",
    eprint = "0712.1144",
    archivePrefix = "arXiv",
    primaryClass = "astro-ph",
    doi = "10.1086/590230",
    journal = "Astrophys. J.",
    volume = "684",
    pages = "870--888",
    year = "2008"
}

@article{McWilliams:2011zs,
    author = "McWilliams, Sean T. and Lang, Ryan N. and Baker, John G. and Thorpe, James Ira",
    title = "{Sky localization of complete inspiral-merger-ringdown signals for nonspinning massive black hole binaries}",
    eprint = "1104.5650",
    archivePrefix = "arXiv",
    primaryClass = "gr-qc",
    doi = "10.1103/PhysRevD.84.064003",
    journal = "Phys. Rev. D",
    volume = "84",
    pages = "064003",
    year = "2011"
}

@Article{Hunter:2007,
  Author    = {Hunter, J. D.},
  Title     = {Matplotlib: A 2D graphics environment},
  Journal   = {Computing in Science \& Engineering},
  Volume    = {9},
  Number    = {3},
  Pages     = {90--95},
  publisher = {IEEE COMPUTER SOC},
  doi       = {10.1109/MCSE.2007.55},
  year      = 2007
}

@ARTICLE{2020SciPy-NMeth,
  author  = {Virtanen, Pauli and Gommers, Ralf and Oliphant, Travis E. and
            Haberland, Matt and Reddy, Tyler and Cournapeau, David and
            Burovski, Evgeni and Peterson, Pearu and Weckesser, Warren and
            Bright, Jonathan and {van der Walt}, St{\'e}fan J. and
            Brett, Matthew and Wilson, Joshua and Millman, K. Jarrod and
            Mayorov, Nikolay and Nelson, Andrew R. J. and Jones, Eric and
            Kern, Robert and Larson, Eric and Carey, C J and
            Polat, {\.I}lhan and Feng, Yu and Moore, Eric W. and
            {VanderPlas}, Jake and Laxalde, Denis and Perktold, Josef and
            Cimrman, Robert and Henriksen, Ian and Quintero, E. A. and
            Harris, Charles R. and Archibald, Anne M. and
            Ribeiro, Ant{\^o}nio H. and Pedregosa, Fabian and
            {van Mulbregt}, Paul and {SciPy 1.0 Contributors}},
  title   = {{{SciPy} 1.0: Fundamental Algorithms for Scientific
            Computing in Python}},
  journal = {Nature Methods},
  year    = {2020},
  volume  = {17},
  pages   = {261--272},
  adsurl  = {https://rdcu.be/b08Wh},
  doi     = {10.1038/s41592-019-0686-2},
}

@Article{harris2020array,
 title         = {Array programming with {NumPy}},
 author        = {Charles R. Harris and K. Jarrod Millman and St{\'{e}}fan J.
                 van der Walt and Ralf Gommers and Pauli Virtanen and David
                 Cournapeau and Eric Wieser and Julian Taylor and Sebastian
                 Berg and Nathaniel J. Smith and Robert Kern and Matti Picus
                 and Stephan Hoyer and Marten H. van Kerkwijk and Matthew
                 Brett and Allan Haldane and Jaime Fern{\'{a}}ndez del
                 R{\'{i}}o and Mark Wiebe and Pearu Peterson and Pierre
                 G{\'{e}}rard-Marchant and Kevin Sheppard and Tyler Reddy and
                 Warren Weckesser and Hameer Abbasi and Christoph Gohlke and
                 Travis E. Oliphant},
 year          = {2020},
 month         = sep,
 journal       = {Nature},
 volume        = {585},
 number        = {7825},
 pages         = {357--362},
 doi           = {10.1038/s41586-020-2649-2},
 publisher     = {Springer Science and Business Media {LLC}},
 url           = {https://doi.org/10.1038/s41586-020-2649-2}
}

@article{DalCanton:2019wsr,
    author = "Dal Canton, Tito and Mangiagli, Alberto and Noble, Scott C. and Schnittman, Jeremy and Ptak, Andrew and Klein, Antoine and Sesana, Alberto and Camp, Jordan",
    title = "{Detectability of modulated X-rays from LISA's supermassive black hole mergers}",
    eprint = "1902.01538",
    archivePrefix = "arXiv",
    primaryClass = "astro-ph.HE",
    doi = "10.3847/1538-4357/ab505a",
    journal = "Astrophys. J.",
    volume = "886",
    pages = "146",
    year = "2019"
}

@article{Marsat:2020rtl,
    author = "Marsat, Sylvain and Baker, John G. and Dal Canton, Tito",
    title = "{Exploring the Bayesian parameter estimation of binary black holes with LISA}",
    eprint = "2003.00357",
    archivePrefix = "arXiv",
    primaryClass = "gr-qc",
    doi = "10.1103/PhysRevD.103.083011",
    journal = "Phys. Rev. D",
    volume = "103",
    number = "8",
    pages = "083011",
    year = "2021"
}

@article{Ruan:2024qch,
    author = "Ruan, Wen-Hong and Guo, Zong-Kuan",
    title = "{Premerger detection of massive black hole binaries using deep learning}",
    eprint = "2402.16282",
    archivePrefix = "arXiv",
    primaryClass = "astro-ph.IM",
    doi = "10.1103/PhysRevD.109.123031",
    journal = "Phys. Rev. D",
    volume = "109",
    number = "12",
    pages = "123031",
    year = "2024"
}

@misc{Colpi:2024xhw,
    author = "Colpi, Monica and others",
    title = "{LISA Definition Study Report}",
    eprint = "2402.07571",
    archivePrefix = "arXiv",
    primaryClass = "astro-ph.CO",
    month = "2",
    year = "2024",
    journal = ""
}

@article{Zackay:2019kkv,
    author = "Zackay, Barak and Venumadhav, Tejaswi and Roulet, Javier and Dai, Liang and Zaldarriaga, Matias",
    title = "{Detecting gravitational waves in data with non-stationary and non-Gaussian noise}",
    eprint = "1908.05644",
    archivePrefix = "arXiv",
    primaryClass = "astro-ph.IM",
    doi = "10.1103/PhysRevD.104.063034",
    journal = "Phys. Rev. D",
    volume = "104",
    number = "6",
    pages = "063034",
    year = "2021"
}

@software{alex_nitz_2024_10473621,
  author       = {Alex Nitz and
                  Ian Harry and
                  Duncan Brown and
                  Christopher M. Biwer and
                  Josh Willis and
                  Tito Dal Canton and
                  Collin Capano and
                  Thomas Dent and
                  Larne Pekowsky and
                  Gareth S Cabourn Davies and
                  Soumi De and
                  Miriam Cabero and
                  Shichao Wu and
                  Andrew R. Williamson and
                  Bernd Machenschalk and
                  Duncan Macleod and
                  Francesco Pannarale and
                  Prayush Kumar and
                  Steven Reyes and
                  dfinstad and
                  Sumit Kumar and
                  M\'arton T\'apai and
                  Leo Singer and
                  Praveen Kumar and
                  veronica-villa and
                  maxtrevor and
                  Bhooshan Uday Varsha Gadre and
                  Sebastian Khan and
                  Stephen Fairhurst and
                  Arthur Tolley},
  title        = {gwastro/pycbc: v2.3.3 release of PyCBC},
  month        = jan,
  year         = 2024,
  publisher    = {Zenodo},
  version      = {v2.3.3},
  doi          = {10.5281/zenodo.10473621},
  url          = {https://doi.org/10.5281/zenodo.10473621},
}

@article{Capano:2021etf,
    author = "Capano, Collin D. and Cabero, Miriam and Westerweck, Julian and Abedi, Jahed and Kastha, Shilpa and Nitz, Alexander H. and Wang, Yi-Fan and Nielsen, Alex B. and Krishnan, Badri",
    title = "{Multimode Quasinormal Spectrum from a Perturbed Black Hole}",
    eprint = "2105.05238",
    archivePrefix = "arXiv",
    primaryClass = "gr-qc",
    doi = "10.1103/PhysRevLett.131.221402",
    journal = "Phys. Rev. Lett.",
    volume = "131",
    number = "22",
    pages = "221402",
    year = "2023"
}

@article{Castelli:2024sdb,
    author = "Castelli, Eleonora and Baghi, Quentin and Baker, John G. and Slutsky, Jacob and Bobin, J{\'e}r{\^o}me and Karnesis, Nikolaos and Petiteau, Antoine and Sauter, Orion and Wass, Peter and Weber, William J.",
    title = "{Extracting gravitational wave signals from LISA data in the presence of artifacts}",
    eprint = "2411.13402",
    archivePrefix = "arXiv",
    primaryClass = "gr-qc",
    doi = "10.1088/1361-6382/adb931",
    journal = "Class. Quant. Grav.",
    volume = "42",
    number = "6",
    pages = "065018",
    year = "2025"
}

@article{Allen:2004gu,
    author = "Allen, Bruce",
    title = "{${\chi}^{2}$ time-frequency discriminator for gravitational wave detection}",
    eprint = "gr-qc/0405045",
    archivePrefix = "arXiv",
    doi = "10.1103/PhysRevD.71.062001",
    journal = "Phys. Rev. D",
    volume = "71",
    pages = "062001",
    year = "2005"
}

@article{Seto:2004ji,
    author = "Seto, Naoki",
    title = "{Annual modulation of the galactic binary confusion noise background and LISA data analysis}",
    eprint = "gr-qc/0403014",
    archivePrefix = "arXiv",
    doi = "10.1103/PhysRevD.69.123005",
    journal = "Phys. Rev. D",
    volume = "69",
    pages = "123005",
    year = "2004"
}

@article{Crowder:2006eu,
    author = "Crowder, Jeff and Cornish, Neil",
    title = "{A Solution to the Galactic Foreground Problem for LISA}",
    eprint = "astro-ph/0611546",
    archivePrefix = "arXiv",
    doi = "10.1103/PhysRevD.75.043008",
    journal = "Phys. Rev. D",
    volume = "75",
    pages = "043008",
    year = "2007"
}

@article{Strub:2024kbe,
    author = {Strub, Stefan H. and Ferraioli, Luigi and Schmelzbach, C{\'e}dric and St{\"a}hler, Simon C. and Giardini, Domenico},
    title = "{Global analysis of LISA data with Galactic binaries and massive black hole binaries}",
    eprint = "2403.15318",
    archivePrefix = "arXiv",
    primaryClass = "gr-qc",
    doi = "10.1103/PhysRevD.110.024005",
    journal = "Phys. Rev. D",
    volume = "110",
    number = "2",
    pages = "024005",
    year = "2024"
}

\end{document}